\documentclass[11pt, a4paper]{article}
\usepackage[utf8]{inputenc}

\usepackage{bbold}
\usepackage{amssymb}
\usepackage{amsmath}
\usepackage{physics}
\usepackage{comment}
\usepackage{xcolor}
\usepackage[margin=35mm]{geometry}
\usepackage{MnSymbol}
\usepackage{mathrsfs, mathtools}

\definecolor{myRed}{rgb}{1.0,0.0,0.0}

\definecolor{myBlue}{rgb}{0.0,0.0,1.0}

\definecolor{myPurple}{rgb}{.4,0.2,0.8}

\definecolor{myGreen}{rgb}{0,0.7,0}

\usepackage{hyperref}
\hypersetup{
    colorlinks=true, 
    linktoc=all,     
    linkcolor=myPurple,  
    citecolor=myGreen,
    menucolor=Black,
    urlcolor=myBlue,
    linktocpage=true,
}

\usepackage{amsthm}

\usepackage{tikz}    
\usetikzlibrary{arrows.meta}
\usetikzlibrary{calc} 
\usetikzlibrary{cd}
\usetikzlibrary{matrix,calc}
\usetikzlibrary{decorations.pathreplacing,positioning}
\usetikzlibrary{decorations.markings,shapes.geometric,shapes.misc}

\def\d{\textup{d}}
\def\g{\mathfrak{g}}
\def\gl{\mathfrak{gl}}
\def\CC{\mathbb{C}}
\def\CP{\mathbb{C}P^1}
\def\RR{\mathbb{R}}
\def\ZZ{\mathbb{Z}}
\def\ii{\mathsf i}

\newcommand{\tensor}[1]{{\mathfrak{#1}}}
\def\1{\tensor{1}}
\def\2{\tensor{2}}
\def\3{\tensor{3}}
\def\4{\tensor{4}}

\begin{document}

\begin{center}
{\LARGE\bf $\bf 3$-dimensional mixed BF theory}\\
\vspace{1.5mm}
{\LARGE\bf and Hitchin's integrable system}\\

\vspace{1.5em}
Beno\^{\i}t Vicedo and Jennifer Winstone

\vspace{1em}
\begingroup\itshape
{\it Department of Mathematics, University of York, York YO10 5DD, U.K.}
\par\endgroup
\vspace{1em}
\begingroup\ttfamily
benoit.vicedo@gmail.com, jennifer.winstone@york.ac.uk
\par\endgroup
\vspace{1.5em}
\end{center}

\begin{abstract}
The affine Gaudin model, associated with an untwisted affine Kac-Moody algebra, is known to arise from a certain gauge fixing of 4-dimensional mixed topological-holomorphic Chern-Simons theory in the Hamiltonian framework. We show that the finite Gaudin model, associated with a finite-dimensional semisimple Lie algebra, or more generally the tamely ramified Hitchin system on an arbitrary Riemann surface, can likewise be obtained from a similar gauge fixing of 3-dimensional mixed BF theory in the Hamiltonian framework.
\end{abstract}




\section{Introduction}

The Heisenberg spin chain can be obtained from $4$-dimensional mixed topological-holomorphic Chern-Simons theory on $\RR^2 \times \CC$ by introducing certain line defects along the topological plane $\RR^2$ for each site of the spin chain \cite{Costello:2013zra, Costello:2013sla, Costello:2017dso, Costello:2018gyb}. This elegant description of the Heisenberg spin chain is ultimately possible because the integrable structure of the latter is underpinned by the \emph{quantum} Yang-Baxter equation
\begin{center}
\raisebox{8mm}{
\begin{minipage}[b]{0.5\linewidth}
\begin{align*}
&R_{\1\2}(z_1, z_2) R_{\1\3}(z_1, z_3) R_{\2\3}(z_2, z_3)\\
&\qquad = R_{\2\3}(z_2, z_3) R_{\1\3}(z_1, z_3) R_{\1\2}(z_1, z_2)
\end{align*}
\end{minipage}
}
\quad
\begin{minipage}[b]{0.4\linewidth}
\begin{tikzpicture}[scale=0.9]
\draw[thick] (-.1,0) node[below=1mm]{$z_2$} -- (-.1,2);
\draw[thick] (1,0) node[below right=1mm and -2mm]{$z_3$} -- (-.5,2);
\draw[thick] (-.5,0) node[below left=1mm and -1mm]{$z_1$} -- (1,2);
\filldraw[thick, fill=gray!60] (.25,1) circle (0.07);
\filldraw[thick, fill=gray!60] (-.1,1.48) circle (0.07);
\filldraw[thick, fill=gray!60] (-.1,.52) circle (0.07);
\end{tikzpicture}
\raisebox{14mm}{=}
\begin{tikzpicture}[scale=0.9]
\draw[thick] (.6,0) node[below=1mm]{$z_2$} -- (.6,2);
\draw[thick] (1,0) node[below right=1mm and -1mm]{$z_3$} -- (-.5,2);
\draw[thick] (-.5,0) node[below left=1mm and -2mm]{$z_1$} -- (1,2);
\filldraw[thick, fill=gray!60] (.25,1) circle (0.07);
\filldraw[thick, fill=gray!60] (.6,1.48) circle (0.07);
\filldraw[thick, fill=gray!60] (.6,.52) circle (0.07);
\end{tikzpicture}
\end{minipage}
\end{center}
which is neatly encoded in the semi-topological nature of the $4$-dimensional Chern-Simons theory. By contrast, the integrability of the (classical and quantum) Gaudin model \cite{Gaudin}, or more generally the Hitchin system \cite{Hitchin}, is underpinned by the \emph{classical} Yang-Baxter equation
\begin{equation} \label{CYBE}
[r_{\1\2}(z_1, z_2), r_{\1\3}(z_1, z_3)] = [r_{\2\3}(z_2, z_3), r_{\1\2}(z_1, z_2)] - [r_{\3\2}(z_3, z_2), r_{\1\3}(z_1, z_3)],
\end{equation}
whose topological origin is less clear.

On the other hand, affine Gaudin models, whose integrability is also underpinned by the classical Yang-Baxter equation \eqref{CYBE}, see \cite{Vicedo:2017cge}, can be obtained \cite{Vicedo:2019dej} from the same $4$-dimensional mixed topological-holomorphic Chern-Simons theory \cite{Costello:2019tri} on $\RR^2 \times \CP$, this time by introducing surface defects along $\RR^2$ placed at the marked points $z_i \in \CC$ of the affine Gaudin model. A natural question is therefore whether the ordinary Gaudin model, associated with a finite-dimensional semisimple Lie algebra $\g$ rather than an affine Kac-Moody algebra, can also be described in a similar way.

\medskip

The purpose of this paper is to show that the tamely ramified Hitchin system on a Riemann surface $C$, in which the Higgs field has simple poles at certain marked points on $C$, can be described using a collection of various line defects in $3$-dimensional mixed (topological-holomorphic) BF theory on $\RR \times C$. In particular, the Gaudin model is obtained as the special case when $C=\CP$. More precisely, we perform a Hamiltonian analysis of the $3$d mixed BF theory, whose fields are a partial connection $1$-form $A$ and a $(1,0)$-form $B$, with suitably chosen line defects. Using the condition $A_{\bar z} = 0$ to fix the gauge invariance, we find that the dynamics on the reduced phase space coincides with that of the tamely ramified Hitchin system. In particular, the $(1,0)$-form $B$ becomes meromorphic and gets identified with the Higgs field. This is completely analogous to the relationship found in \cite{Vicedo:2019dej} between $4$d mixed topological-holomorphic Chern-Simons theory on $\Sigma \times \CP$, with the cylinder $\Sigma = \RR \times S^1$ , and the affine Gaudin model. In other words, our analysis shows that $3$d mixed BF theory is to the Gaudin model what $4$d Chern-Simons theory is to the affine Gaudin model.

\medskip

The plan of the paper is as follows.

In \S\ref{sec: Lag analysis} we recall the definition of $3$-dimensional mixed topological-holomorphic BF theory on $\RR \times C$ for some Riemann surface $C$, depending on a $\g$-valued $(1,0)$-form $B$ and a $\g$-valued connection $1$-form $A$. In \S\ref{sec: defects} we introduce two types of line defects, which we refer to as type $A$ and $B$, respectively. Type $A$ defects ensure that in the gauge $A_{\bar z} = 0$ the field $B$ is meromorphic on-shell, with poles at the location of the defects. This field then gets identified with the Lax matrix $L$ of the Gaudin model, or the Higgs field $\Phi$ of the tamely ramified Hitchin system. Adding the type $B$ defect ensures that (the negative of) the time component of the connection $1$-form $A$ gets correctly identified with the matrix $M$ in the Lax equation $\partial_t L = [M, L]$. We also derive the $1$d action for the Gaudin model along the lines of \cite{Delduc:2019whp, Caudrelier:2020xtn}.

In \S\ref{sec: Ham analysis} we perform a Hamiltonian analysis of $3$-dimensional mixed topological-holomorphic BF theory with the two types of defects introduced in \S\ref{sec: defects}. There is a first class constraint $\widehat \mu$ generating the gauge symmetry of the theory. In the absence of type $A$ defects, $\widehat{\mu}$ coincides with the moment map $\mu = \bar \partial_A \Phi$ of the Hitchin system, after identifying the Higgs field $\Phi$ with the $(1,0)$-form $B$ of the $3$-dimensional mixed topological-holomorphic BF theory. Upon imposing the condition $A_{\bar z} = 0$ to fix the gauge symmetry generated by $\widehat{\mu}$, the component $B_z$ of the $(1,0)$-form $B$ becomes meromorphic with poles at the location $z_i$ of the type $A$ line defects and is shown to satisfy the Lax algebra with respect to the Dirac bracket. Morever, the Hamiltonian on the reduced phrase space coincides with that of the Hitchin system.

We end the paper with a brief discussion of possible future directions in \S\ref{sec: discussion}.

\paragraph{Acknowledgements:} This paper was motivated by a talk given by E. Rabinovich at the online workshop ``A Gauge Summer with BV’’ in June 2020, where the relationship between the factorisation algebra of quantum observables of 3d mixed BF theory and the derived center of the Kac-Moody vertex algebra at the critical level, and in particular its relation to the Gaudin algebra, was first proposed. The first author is very grateful to O. Gwilliam, E. Rabinovich and B. R. Williams for many interesting and inspiring discussions in relation to this talk. We are also grateful to S. Lacroix for fruitful discussions. This work was supported by the Engineering and Physical Research Council (Grant number EP/R513386/1).

\section{\texorpdfstring{$\bf 3$}{3}d mixed BF theory on \texorpdfstring{$\RR \times C$}{RxC}} \label{sec: Lag analysis}

Let $G$ be a semisimple Lie group over $\CC$, with Lie algebra $\g$ and fix a non-degenerate invariant symmetric bilinear form $\langle \cdot, \cdot \rangle : \g \times \g \to \CC$ on $\g$. Let $C$ be a Riemann surface.

We shall consider the $3$-dimensional classical mixed topological-holomorphic BF theory on $\RR \times C$, or \emph{$3$d mixed BF theory} for short -- see \emph{e.g.} \cite{GW, GRW, RaPhD} where the theory is discussed using the BV formalism. The field content of this theory consists of a $\g$-valued $(1,0)$-form $B$ on $\RR \times C$, where the bigrading $(p,q)$ corresponds to the one induced by the complex structure on $C$, together with a $\g$-valued connection $1$-form $A$ on $\RR \times C$. We denote the curvature of the latter as $F(A) = \d A+ \frac 12 [A,A]$. The action of $3$d mixed BF theory is then given by
\begin{equation} \label{BF action}
S_{\rm 3d}[A,B] = \frac{1}{2\pi \ii} \int_{\RR \times C} \langle B, F(A) \rangle.
\end{equation}

\subsection{Gauge invariance}

The $3$d mixed BF action \eqref{BF action} is trivially invariant under gauge transformations of the form $A \to A + \chi$ for any $\g$-valued $(1,0)$-form $\chi$ on $\RR \times C$. Indeed, $\chi$ drops out from the action since $B$ is a $(1,0)$-form. We can fix this invariance by requiring that $A$ has no $(1,0)$-component so that it locally takes the form $A = A_{\bar z} \d \bar z + A_t \d t$ for some local coordinate $t$ on $\RR$ and a local holomorphic coordinate $z$ on $C$. From now on we will always take $A$ to be a partial connection of this form.

More interestingly, the action \eqref{BF action} is invariant under the action of any $G$-valued function $g$ on $\RR \times C$ acting by gauge transformations on the connection $1$-form $A$ and by conjugation on the field $B$, namely
\begin{subequations} \label{gauge transf A B}
\begin{align}
A &\longmapsto \null^g A \coloneqq - \bar\partial g g^{-1} - \d_\RR g g^{-1} + g A g^{-1},\\
B &\longmapsto g B g^{-1},
\end{align}
\end{subequations}
where $\bar\partial$ is the Dolbeault differential on $C$, \emph{i.e.} the $(0,1)$-component of the de Rham differential $\d_C = \partial + \bar \partial$ on $C$, and $\d_\RR$ denotes the de Rham differential on $\RR$.
Indeed, under such a transformation the curvature $2$-form $F(A)$ transforms by conjugation $F(\null^g A) = g F(A) g^{-1}$ and so the invariance of the action follows from the adjoint $G$-invariance of the bilinear form $\langle \cdot, \cdot \rangle$.

\subsection{Equations of motion} \label{sec: eoms}

To derive the equations of motion we consider the variations $B \to B + \epsilon$ and $A \to A + \eta$ by an arbitrary $(1,0)$-form $\epsilon$ and $1$-form $\eta$ on $\RR \times C$.
Varying the action we find
\begin{align*}
\delta S_{\rm 3d}[A,B] &\coloneqq S_{\rm 3d}[A+\eta, B+\epsilon] - S_{\rm 3d}[A,B]\\
&\, = \frac{1}{2 \pi \ii} \int_{\RR \times C}
\big( \langle \epsilon, F(A) \rangle + \langle B, \d\eta + \tfrac 12 [A,\eta] + \tfrac 12 [\eta,A] \rangle + O(\eta^2)\big)\\
&\, = \frac{1}{2 \pi \ii} \int_{\RR \times C}
\big( \langle \epsilon, F(A) \rangle + \langle B, \d\eta + [A,\eta] \rangle + O(\eta^2) \big)\\
&\, = \frac{1}{2 \pi \ii} \int_{\RR \times C}
\big( \langle \epsilon, F(A) \rangle + \langle \d B +[B,A], \eta \rangle + O(\eta^2) \big),
\end{align*}
where in the third equality we have used the fact that $A$ and $\eta$ are both $\g$-valued $1$-forms, so that $[\eta, A] = [A, \eta]$. In the last equality we used Stokes's theorem, noting that $\langle B, \d \eta\rangle = \langle \d B, \eta \rangle - \d \langle B, \eta \rangle$, and the adjoint invariance of the bilinear form.

The equation of motion for $B$ is therefore $F(A) = 0$, or explicitly
\begin{subequations} \label{eom}
\begin{equation} \label{eom B}
\bar\partial A + \d_\RR A + \tfrac 12 [A,A] = 0,
\end{equation}
while the equation of motion for $A$ reads
\begin{equation} \label{eom A}
\bar\partial B + \d_\RR B + [B,A] = 0.
\end{equation}
\end{subequations}
Letting $z$ be a local holomorphic coordinate on $C$ and $t$ a global coordinate on $\RR$, and writing the two fields in components as $B = B_z \d z$ and $A = A_{\bar z} \d \bar z + A_t \d t$, we can write the equations of motion \eqref{eom} more explicitly in components as
\begin{subequations} \label{eom coord}
\begin{align}
\label{eom coord a} \partial_{\bar z} A_t - \partial_t A_{\bar z} &= [A_t, A_{\bar z}],\\
\label{eom coord b} \partial_{\bar z} B_z &= [B_z, A_{\bar z}],\\
\label{eom coord c} \partial_t B_z &= [-A_t, B_z].
\end{align}
\end{subequations}

The first key observation to make here is that the equation of motion \eqref{eom coord c} is very reminiscent of the Lax equation
\begin{equation} \label{Lax eq}
\partial_t L = [M, L].    
\end{equation}
However, to make this superficial resemblance more precise we would need $B_z$ and $- A_t$ to both be holomorphic (or more generally meromorphic) in order to identify them with the Lax pair $L$ and $M$ of an integrable system.

The second observation, based on the other two equations of motion \eqref{eom coord a} and \eqref{eom coord b}, is that this can be achieved by working in the gauge where $A_{\bar z} = 0$. Indeed, in this gauge the two equations \eqref{eom coord a} and \eqref{eom coord b} reduce to $\partial_{\bar z} A_t = 0$ and $\partial_{\bar z} B_z = 0$, respectively, which express the fact that $A_t$ and $B_z$ are both holomorphic on $C$.

\subsection{Introducing line defects} \label{sec: defects}

In the Lax equation \eqref{Lax eq} of an integrable system, however, $L$ and $M$ are more generally $\g$-valued \emph{meromorphic} functions with poles at certain marked points. This is, in fact, necessary if $C$ has genus zero, \emph{i.e.} when $C = \CP$. Moreover, as it stands there is no relation between $B_z$ and $- A_t$ in \eqref{eom coord c}, while in \eqref{Lax eq} the matrix $M$ is typically built out of the Lax matrix $L$. We can fix both of these issues by introducing two different types of line defects in the action \eqref{BF action}. We will refer to these as type $A$ and type $B$ line defects, since these will depend on the fields $A$ and $B$, respectively.

\subsubsection{Type \texorpdfstring{$A$}{A} line defects} \label{sec: A defect}

The Lax pair of the Gaudin model is formed of two $\g$-valued meromorphic functions $L$ and $M$ on $\CP$ with $L$ having poles at certain marked points $z_i \in \CC$ for $i = 1, \ldots, N$. In order to view $B_z$ and $- A_t$ as such a Lax pair, but working on a more general Riemann surface $C$, we would like them to be meromorphic instead of holomorphic, with $B_z$ having poles at certain marked points $z_i \in C$. To this end, we pick and fix elements $u_i \in \g$ and introduce $G$-valued fields $h_i$ on $\RR$ for $i=1,\ldots, N$. Following \cite{Costello:2019tri}, see also \cite{Caudrelier:2020xtn}, we add to the action \eqref{BF action} the following sum of line defects
\begin{equation} \label{A defect}
S_{A-{\rm def}}\big[ A, \{h_i\}_{i=1}^N \big] = - \sum_{i=1}^N \int_{\mathbb R \times \{z_i\}} \big\langle u_i, h_i^{-1}(\d_\RR + \iota_{z_i}^\ast A) h_i \big\rangle
\end{equation}
where $\iota_{z_i} : \RR \times \{z_i\} \hookrightarrow \RR \times C$ is the embedding of the line defect at $z_i$ into the total space. In particular, the pullback $\iota_{z_i}^\ast A$ is just the evaluation of the component $A_t \d t$ at the point $z_i \in C$ so that we can rewrite the defect action \eqref{A defect} more explicitly as
\begin{equation*}
S_{A-{\rm def}}\big[ A, \{h_i\}_{i=1}^N \big] = - \sum_{i=1}^N \int_\RR \big\langle u_i, h_i^{-1}\big(\partial_t + A_t(z_i)\big) h_i \big\rangle \d t.
\end{equation*}

In order to maintain the gauge invariance of the action \eqref{BF action} after adding \eqref{A defect} to it, we should require that the latter is itself gauge invariant. This can easily be achieved by supplementing the gauge transformations \eqref{gauge transf A B} of the fields $A$ and $B$ by the transformation
\begin{equation} \label{gauge transf phi}
h_i \longmapsto g h_i
\end{equation}
for the $G$-valued fields $h_i$, $i = 1, \ldots, N$.

Consider now the extended action
\begin{equation} \label{BF action + A}
\widetilde S\big[ A,B, \{h_i\}_{i=1}^N \big] \coloneqq S_{\rm 3d}[A,B] + S_{A-{\rm def}}\big[ A, \{h_i\}_{i=1}^N \big].
\end{equation}
Since the defect action \eqref{A defect} does not depend on $B$, the equations of motion \eqref{eom B} for $B$ are unchanged. On the other hand, the equation of motion \eqref{eom coord b} for $A$ in a local holomorphic coordinate $z$ on an open neighbourhood $U$ of the point $z_i$ is now replaced by
\begin{equation} \label{eom coord b def}
\partial_{\bar z} B_z = [B_z, A_{\bar z}] - 2\pi \ii \, \widehat{u}_i \delta_{z z_i},
\end{equation}
where we introduced $\widehat{u}_i \coloneqq h_i u_i h_i^{-1}$ for each $i = 1,\ldots, N$ and $\delta_{z z_i}$ denotes the Dirac $\delta$-distribution at the marked point $z_i$ with the property that
\begin{equation*}
\int_U f(z) \delta_{z z_i} \d z \wedge \d \bar z = f(z_i)
\end{equation*}
for any function $f : U \to \CC$ on the neighbourhood $U \subset C$ of $z_i$ equipped with the local holomorphic coordinate $z$.

In the gauge $A_{\bar z} = 0$, the modified equation of motion \eqref{eom coord b def} reads
\begin{equation} \label{Bz mero}
\partial_{\bar z} B_z = - 2 \pi \ii \, \widehat{u}_i \delta_{z z_i}.
\end{equation}
Using the fact that $\partial_{\bar z} (z-z_i)^{-1} = - 2 \pi \ii \delta_{zz_i}$ we may rewrite this equation as
\begin{equation*}
\partial_{\bar z} \bigg( B_z - \frac{\widehat{u}_i}{z-z_i} \bigg) = 0
\end{equation*}
which tells us that $B_z$ has a simple pole at $z_i$ with residue $\widehat u_i$ there, \emph{i.e.}
\begin{equation} \label{B mero Hitchin}
B = \frac{\widehat{u}_i}{z-z_i} \d z + O(1)
\end{equation}
where $O(1)$ denotes terms which are holomorphic at the point $z_i$.

When $C = \CP$, corresponding to the Gaudin model, if we fix a global coordinate $z$ on $\CC \subset \CP$ and require $B$ to have a simple pole also at infinity then we can explicitly write $B$ as the $\g$-valued meromorphic $(1,0)$-form
\begin{equation} \label{B mero}
B = \sum_{i=1}^N \frac{\widehat{u}_i}{z - z_i} \d z.
\end{equation}

By varying the action \eqref{BF action + A} with respect to $h_i \to e^{\epsilon_i} h_i$ for some $\g$-valued function $\epsilon_i$ on $\RR$ we find $N$ further equations of motion
\begin{equation} \label{hat u i eom}
\partial_t \widehat{u}_i = [ - A_t(z_i), \widehat{u}_i ]
\end{equation}
for $i = 1, \ldots, N$. But given the meromorphic behaviour \eqref{B mero Hitchin} of the $(1,0)$-form $B$ at each of the marked points $z_i$, these are merely consequences of the equation of motion \eqref{eom coord c} given by taking the residue at each $z_i$, assuming that $A_t$ is regular at $z_i$, as will be the case in \S\ref{sec: B defect}.

\subsubsection{Type \texorpdfstring{$B$}{B} line defects} \label{sec: B defect}

The type $A$ line defects introduced in \S\ref{sec: A defect} ensured that $B_z$ is no longer holomorphic in the gauge $A_{\bar z} = 0$ but rather meromorphic with poles at certain marked points $z_i \in C$. The type $B$ line defects will have a similar effect on the field $A_t$. However, since $- A_t$ is meant to play the role of $M$ in the Lax pair \eqref{Lax eq}, we want it to be built out of $B_z$, which plays the role of the Lax matrix $L$.

Let $P : \g \to \CC$ be a $G$-invariant polynomial on $\g$ and fix a point $w \in C$ distinct from the marked points $z_i \in C$ for $i=1,\ldots, N$ at which the type $A$ line defects were inserted in \S\ref{sec: A defect}.
We consider the following line defect
\begin{equation} \label{B defect}
S_{B-{\rm def}}[B] = - \int_{\mathbb R \times \{w\}} P(B_z) \d t = - \int_\RR P\big( B_z(w) \big) \d t
\end{equation}
where $z$ is a local holomorphic coordinate around the point $w \in C$ and, writing $B = B_z \d z$ in this coordinate, $B_z(w)$ denotes the evaluation of $B_z$ at the point $w$.

The $G$-invariance of the polynomial $P$ ensures that the action \eqref{B defect} is gauge invariant. Therefore, adding it to the gauge invariant action \eqref{BF action + A} obtained so far, we obtain the full gauge invariant action
\begin{equation} \label{BF action + A + B}
S\big[ A,B, \{h_i\}_{i=1}^N \big] \coloneqq S_{\rm 3d}[A,B] + S_{A-{\rm def}}\big[ A, \{h_i\}_{i=1}^N \big] + S_{B-{\rm def}}[B].
\end{equation}

Since the defect term \eqref{B defect} only depends on $B$, it will not modify the equations of motion for $A$. Only the equation of motion for $B$, namely \eqref{eom coord a} which has so far remained unchanged, will be modified. To derive it we note that the variation of the defect action \eqref{B defect}, under the variation $B \to B + \epsilon$ considered in \S\ref{sec: eoms} with $\epsilon = \epsilon_z \d z$ in the local holomorphic coordinate $z$, reads
\begin{align*}
\delta S_{B-{\rm def}}[B] &\coloneqq S_{B-{\rm def}}[B+\epsilon] - S_{B-{\rm def}}[B]\\
&= - \int_\RR \Big( P\big( B_z(w) + \epsilon_z(w) \big) - P\big( B_z(w) \big) \Big) \d t\\
&= - \int_\RR \Big( \big\langle P'(B_z(w)), \epsilon_z(w) \big\rangle + O\big( \epsilon_z(w)^2 \big) \Big) \d t
\end{align*}
where in the third line we introduced the element $P'(B_z(w)) \in \g$ such that the linear map $\langle P'(B_z(w)), \cdot \rangle : \g \to \CC$ is the derivative of $P: \g \to \CC$ at $B_z(w)$ and kept only the terms linear in $\epsilon_z(w)$. It follows that \eqref{eom coord a} is now replaced by
\begin{equation} \label{eom coord a def}
\partial_{\bar z} A_t - \partial_t A_{\bar z} = [A_t, A_{\bar z}] + 2 \pi \ii \, P'\big( B_z(w) \big) \delta_{zw}.
\end{equation}
In the gauge $A_{\bar z} = 0$ this simplifies to
\begin{equation} \label{bar z At}
\partial_{\bar z} A_t = 2 \pi \ii \, P'\big( B_z(w) \big) \delta_{zw}
\end{equation}
or in other words,
\begin{equation*}
\partial_{\bar z} \bigg( A_t + \frac{P'\big( B_z(w) \big)}{z - w} \bigg) = 0.
\end{equation*}

In the case $C = \CP$ this tells us that the expression in brackets is a constant. Taking this contant to be zero we therefore obtain
\begin{equation} \label{At Gaudin}
- A_t(z) = \frac{P'\big( B_z(w) \big)}{z - w},
\end{equation}
which coincides with the usual expression for $M = - A_t$ in terms of $L = B_z$, see for instance \cite[(3.33)]{BBT} in the case when $\g = \gl_r$ and the polynomial $P : \gl_r \to \CC$ is given by $X \mapsto \textup{tr}(X^n)$ for some $n \in \ZZ_{\geq 1}$. Indeed, in this case we have $P'(X) = n X^{n-1}$ for any $X \in \gl_r$ so that \eqref{At Gaudin} becomes
\begin{equation} \label{At Gaudin glr}
- A_t(z) = n \frac{B_z(w)^{n-1}}{z - w}.
\end{equation}
In connection with the Hamiltonian analysis to be performed in \S\ref{sec: Ham analysis} below, where the classical $r$-matrix $r_{\1\2}(z,w) = \frac{C_{\1\2}}{w-z}$ will be introduced in \eqref{r-matrix}, note that we can rewrite \eqref{At Gaudin glr} in the more recognisable form
\begin{equation*}
- A_t(z) = - n \, \textup{tr}_\2 \big( r_{\1\2}(z,w) B_z(w)_\2^{n-1} \big).
\end{equation*}

Substituting the expression \eqref{At Gaudin} for $A_t$ into the equation of motion \eqref{eom coord c} we obtain the desired Lax equation
\begin{equation} \label{Lax equation Bz}
\partial_t B_z(z) = \bigg[ \frac{P'\big( B_z(w) \big)}{z - w}, B_z(z) \bigg],
\end{equation}
where we have explicitly written the dependence of $B_z$ on the spectral parameters.
We thus expect from the general theory of integrable systems, see for instance the Proposition \cite[p.47]{BBT}, that the time coordinate $t$ along the topological direction of the $3$-dimensional space $\RR \times C$ is identified, through the introduction of the type $B$ defect \eqref{B defect}, with the time induced by the Hamiltonian
\begin{equation} \label{Ham Hitchin}
H^P_w \coloneqq P(B_z(w)).
\end{equation}
To confirm this we will move to the Hamiltonian formalism in \S\ref{sec: Ham analysis} below.

\subsection{Unifying \texorpdfstring{$\mathbf 1$}{1}-dimensional action}

We have now shown that the gauge fixed equations of motion for $3$d mixed BF theory in the presence of type $A$ and $B$ defects correspond exactly to the Lax equation \eqref{Lax equation Bz} of the Gaudin model with Lax matrix $L(z) = B_z(z)$ given by \eqref{B mero}, where the residues $\widehat{u}_i = h_i u_i h_i^{-1}$ are coadjoint orbits through the fixed elements $u_i \in \g$ and parametrised by the dynamical $G$-valued variables $h_i \in G$.

At this stage it is therefore natural to proceed along the lines of \cite{Delduc:2019whp}, where a unifying $2$d action for integrable field theories of affine Gaudin type was derived from the $4$d Chern-Simons action of \cite{Costello:2019tri}. In a similar spirit, in the present context we would like to obtain a $1$d action for the Gaudin model with Lax matrix \eqref{B mero} starting from the $3$d mixed BF theory with both type $A$ and type $B$ defects. In fact, the procedure followed in \cite{Caudrelier:2020xtn} is closer in spirit to the present case since we do not have to deal with the presence of a meromorphic $1$-form $\omega$ having zeroes, as in the $4$d Chern-Simons action considered in \cite{Delduc:2019whp}.

Following \cite{Caudrelier:2020xtn}, we will therefore substitute the solutions to the equations of motion \eqref{eom coord b def} and \eqref{eom coord a def} (but crucially not \eqref{eom coord c}) in the gauge $A_{\bar z} = 0$, namely \eqref{B mero} and \eqref{At Gaudin} respectively, into the full action \eqref{BF action + A + B}. We will do this for the three pieces in the action separately. For the bulk action \eqref{BF action} we find
\begin{align*}
S_{\rm 3d}[A,B] 
&= \frac{1}{2 \pi \ii} \int_{\RR \times \CC} \langle B_z, \partial_{\bar z} A_t - \partial_t A_{\bar z} - [A_t, A_{\bar z}] \rangle \d z \wedge \d\bar z \wedge \d t\\
&= \frac{1}{2 \pi \ii} \int_{\RR \times \CC} \langle B_z, \partial_{\bar z} A_t \rangle \d z \wedge \d\bar z \wedge \d t\\
&= \int_{\RR} \big\langle L(w), P'\big( L(w) \big) \big\rangle \d t
\end{align*}
where in the second equality we used the gauge $A_{\bar z} = 0$. In the last equality we used the fact that $B_z$ is identified with the Lax matrix $L$ together with the identity \eqref{bar z At}, and then performed the integral over $\CC$ using the presence of the $\delta$-function.

For the type $A$ defect action \eqref{A defect} we have
\begin{align*}
S_{A-{\rm def}}\big[ A, \{h_i\}_{i=1}^N \big]
&= - \sum_{i=1}^N \int_{\RR} \langle u_i, h_i^{-1}\partial_t h_i \rangle \d t - \sum_{i=1}^N \int_{\RR} \langle \widehat u_i, A_t(z_i) \rangle \d t\\
&= - \sum_{i=1}^N \int_{\RR} \langle u_i, h_i^{-1}\partial_t h_i \rangle \d t - \sum_{i=1}^N \int_{\RR} \bigg\langle \widehat u_i, \frac{P'(L(w))}{w - z_i} \bigg\rangle \d t\\
&= - \sum_{i=1}^N \int_{\RR} \langle u_i, h_i^{-1}\partial_t h_i \rangle \d t - \int_{\RR} \big\langle L(w), P'\big( L(w)\big) \big\rangle \d t,
\end{align*}
where in the second equality we used \eqref{At Gaudin} evaluated at $z = z_i$ and in the last line we recognised the sum over $i$ in the second term as the expression for the Lax matrix $L(w) = B_z(w)$ in \eqref{B mero}. Note that the second term on the right hand side exactly cancels the expression found above for the bulk action $S_{\rm 3d}[A,B]$.

Finally, the type $B$ defect action \eqref{B defect} is simply $S_{B-{\rm def}}[B] = - \int_\RR H^P_w \d t$ using the expression \eqref{Ham Hitchin} for the Hamiltonian alluded to in \S\ref{sec: B defect} and to be confirmed in \S\ref{sec: Ham analysis}.
Putting all the above together, we deduce that the full action \eqref{BF action + A + B} reduces to the simple form
\begin{equation} \label{unifying 1d}
S_{\rm 1d}\big[ \{h_i\}_{i=1}^N \big] = - \sum_{i=1}^N \int_{\RR} \langle u_i, h_i^{-1}\partial_t h_i \rangle \d t - \int_{\RR} H^P_w \d t,
\end{equation}
where we have suppressed the dependence on the fields $A$ and $B$ since these have now been expressed in terms of the dynamical variables $h_i \in G$ and the fixed elements $u_i \in \g$ for $i = 1, \ldots, N$. We recognise \eqref{unifying 1d} as the first order action
\begin{equation*}
S\big[ \{h_i\}_{i=1}^N \big] = \sum_{i=1}^N \int_{\RR} \langle X_i, h_i^{-1}\partial_t h_i \rangle \d t - \int_{\RR} H^P_w \d t,
\end{equation*}
associated with the Hamiltonian $H^P_w$ in \eqref{Ham Hitchin} but where the conjugate momentum $X_i \in \g$ of $h_i \in G$ has been fixed to the constant element $X_i = - u_i$. This is consistent with the Hamiltonian analysis to be performed in the next section. Namely, we will find in \S\ref{sec: defect can var} that there is a primary constraint $X_i + u_i \approx 0$ on the conjugate momentum $X_i \in \g$ of the dynamical variable $h_i \in G$.

We can check directly that the equations of motion of the $1$d action \eqref{unifying 1d} are given by \eqref{hat u i eom}, with $A_t$ as in \eqref{At Gaudin}, by varying it with respect to $h_i \to e^{\epsilon_i} h_i$ for some arbitrary $\g$-valued variable $\epsilon_i$. Under this variation, the Lax matrix $L(w)$ transforms to
\begin{equation*}
\sum_{i=1}^N \frac{e^{\epsilon_i}\widehat{u}_i e^{-\epsilon_i}}{w-z_i} = L(w)  +\sum_{i=1}^N\frac{\comm{\epsilon_i}{\widehat{u}_i}}{w-z_i} + O(\epsilon_i^2).
\end{equation*}
Hence, using the explicit expression $H= P(L(w))$ for the Hamiltonian, the variation of the action is given by
\begin{align*}
\delta S_{\rm1d} &\coloneqq S_{\rm 1d}[\{e^{\epsilon_i} h_i\}_{i=1}^N] -  S_{\rm 1d}[\{h_i\}_{i=1}^N]\\
&= - \sum_{i=1}^N \int_{\RR} \big\langle u_i, h_i^{-1}e^{-\epsilon_i}\partial_t (e^{\epsilon_i}h_i) - h_i^{-1}\partial_t h_i \big\rangle \d t\\
&\qquad\qquad - \int_{\RR} \bigg( P\bigg(L(w) +\sum_{i=1}^N\frac{\comm{\epsilon_i}{\widehat{u}_i}}{w-z_i}\bigg) - P\big(L(w)\big) \bigg) \d t\\
&= - \sum_{i=1}^N \int_{\RR} \bigg( \langle \widehat{u}_i, \partial_t \epsilon_i \rangle + \bigg\langle P'\big(L(w)\big), \frac{\comm{\epsilon_i} {\widehat{u}_i}}{w-z_i} \bigg\rangle + O(\epsilon_i^2) \bigg)\d t\\
&= \sum_{i=1}^N \int_{\RR} \bigg( \bigg\langle \partial_t \widehat{u}_i - \comm{\frac{P'(L(w))}{z_i - w}}{\widehat{u}_i}, \epsilon_i \bigg\rangle + O(\epsilon_i^2) \bigg) \d t, 
\end{align*}
where in the last equality we have used Stokes's theorem and the adjoint invariance of the bilinear form.
The \(N\) equations of motion for the \(h_i\) are therefore
\[
    \partial_t \widehat{u}_i = \comm{\frac{P'(L(w))}{z_i -w}}{\widehat{u}_i},
\]
and using \eqref{At Gaudin}, we do indeed recover the equations of motion \eqref{hat u i eom} found previously from adding in type $A$ defects in \S\ref{sec: A defect}.

\section{Hamiltonian analysis} \label{sec: Ham analysis}

Throughout this section we shall work in some local coordinate $z$ on some open subset of $C$. Our starting point is the Lagrangian density of the action \eqref{BF action + A + B} written in terms of the components of the $\g$-valued bulk fields $A = A_{\bar z} \d \bar z + A_t \d t$ and $B = B_z \d z$ and in terms of the $G$-valued defect variables $h_i$ for $i = 1, \ldots, N$ which we introduced at the type $A$ defects in \S\ref{sec: A defect}, namely 
\begin{align} \label{Lagrangian density}
\mathcal{L}\big(A,B, \{h_i\}_{i=1}^N \big)
&= \frac{1}{2 \pi \ii}\big\langle B_z, \partial_{\Bar z} A_t - \partial_t A_{\Bar z} + \comm{A_{\Bar z}}{A_t} \big\rangle \notag\\
&\qquad\qquad - \sum_{i=1}^N
\big\langle u_i, h_i^{-1}(\partial_t + A_t)h_i \big\rangle \delta_{z z_i} - P(B_z(w)) \delta_{zw}.
\end{align}

\subsection{Conjugate momenta and primary constraints} \label{sec: primary constraints}

To move to the Hamiltonian formalism we first determine the conjugate momenta of the bulk fields $A_{\bar z}$, $A_t$ and $B_z$ and the defect variables $h_i$. We shall find various primary constraints, some of which will be second class. We shall impose the latter strongly at this stage by introducing a corresponding Dirac bracket. To alleviate the notation, all Dirac brackets computed in this section will ultimately be renamed simply as $\{ \cdot, \cdot\}$ before moving on to \S\ref{sec: Ham and second constraints} where we work out the secondary constraints.

We will begin in \S\ref{sec: bulk can var} by considering the conjugate momenta of the bulk fields $A_{\bar z}$, $A_t$ and $B_z$, as the conjugate momenta to the $G$-valued defect variables $h_i$ will need to be handled with more care, as discussed in \S\ref{sec: defect can var} below.

\subsubsection{Bulk canonical fields} \label{sec: bulk can var}

The conjugate momenta to the $\g$-valued bulk fields $A_{\bar z}$, $A_t$ and $B_z$ are the $\g$-valued fields given respectively by
\begin{equation*}
\Pi_t  =\frac{\partial \mathcal L}{\partial(\partial_t A_t)} = 0,
\quad
\Pi_{\bar z}  =\frac{\partial \mathcal L}{\partial(\partial_t A_{\bar z})} =  - \frac{1}{2 \pi \ii}B_z,
\quad
P_z  =\frac{\partial \mathcal L}{\partial(\partial_t B_z)} = 0,
\end{equation*}
which satisfy the canonical Poisson bracket relations
\begin{align*}
\{\Pi_{t\1}(z), A_{t\2}(z')\} &= C_{\1\2}\delta_{zz'},\\
\{\Pi_{\bar{z}\1}(z), A_{\bar{z}\2}(z')\} &= C_{\1\2}\delta_{zz'},\\
\{P_{z\1}(z),B_{z\2}(z') \} &=  C_{\1\2}\delta_{zz'}.
\end{align*}

We have three primary constraints associated with the bulk fields, namely
\begin{equation} \label{bulk constraints}
\Pi_t \approx 0 , 
\qquad
\mathcal C_{z} \coloneqq B_z  + 2 \pi \ii \Pi_{\bar z} \approx 0,
\qquad
P_z \approx 0.
\end{equation}
The first is clearly first class and the latter two are second class with Poisson bracket
\begin{gather*}
   \{P_{z\1}(z), \mathcal C_{z\2}(z')\} = C_{\1\2} \delta_{z z'}.
\end{gather*}
We set these both strongly to zero immediately by introducing the corresponding Dirac bracket, with respect to which we still have the same relations between the remaining fields, \emph{i.e.}
\begin{subequations} \label{DB for Cz Pz}
\begin{align}
\label{DB for Cz Pz a} \{\Pi_{t\1}(z), A_{t\2}(z')\} &=  C_{\1\2}\delta_{z z'},\\
\label{DB for Cz Pz b} \{\Pi_{\bar z \1}(z), A_{\bar z \2}(z')\} &=  C_{\1\2}\delta_{z z'},
\end{align}
\end{subequations}
and hence, by an abuse of notation, we will continue to denote this Diract bracket as \(\{\cdot, \cdot\}\).

\subsubsection{Defect canonical variables} \label{sec: defect can var}

We have yet to find the conjugate momenta to the $G$-valued variables $h_i$, $i = 1, \ldots, N$ introduced at the type $A$ defects. This can be done by working in local coordinates $\phi^\alpha$ on the group $G$ where $\alpha$ ranges from 1 to $\dim G$, the dimension of $G$. We refer the reader, for instance, to \cite[\S3.1.2]{Lacroix:2018njs} for details. Each variable $h_i \in G$ can then be described locally in terms of the $\dim G$ variables $\phi^\alpha_i \coloneqq \phi^\alpha(h_i)$.

The relevant part of the Lagrangian in finding the conjugate momenta is
\begin{equation*}
- \langle u_i, h_i^{-1} \partial_t h_i \rangle = - \langle u_i, \partial_t \phi^\alpha_i h_i^{-1}\partial_\alpha h_i \rangle,
\end{equation*}
which in the second expression we have rewritten in terms of the local coordinates $\phi^\alpha_i$, where $\partial_\alpha$ denotes the partial derivative with respect to the coordinate $\phi^\alpha$. 
The corresponding conjugate momenta are therefore given by 
\begin{equation} \label{pi i alpha}
\pi_{i, \alpha} = \frac{\partial \mathcal L}{ \partial (\partial_t \phi^\alpha_i)} = - \langle u_i, h_i^{-1} \partial_\alpha h_i \rangle,
\end{equation}
and these have the usual canonical Poisson bracket relations
\begin{equation*}
    \{\phi^\alpha_i,\phi^\beta_j\} = 0,
    \qquad
    \{\pi_{i, \alpha},\pi_{j, \beta} \} = 0,
    \qquad
    \{\pi_{i, \alpha},\phi^\beta_j\} = \delta^\beta_\alpha \delta_{ij}.
\end{equation*}

To return to a coordinate free description of the phase space, we define a matrix $L^a_{\;\; \alpha}$ for some fixed basis $\{I_a\}$ of $\g$ such that 
\begin{equation} \label{L a alpha def}
h_i^{-1}\partial_\alpha h_i = L^a_{\;\; \alpha} I_a.
\end{equation}
This $L^a_{\;\; \alpha}$ is invertable and we denote the inverse as $L^\alpha_{\;\; a}$ following the conventions of \cite[\S3.1.2]{Lacroix:2018njs}. We can then introduce a coordinate-free $\g$-valued variable $X_i$ which encodes the conjugate momentum $\pi_{i, \alpha}$ as
\begin{equation} \label{Xi def}
X_i \coloneqq L^\alpha_{\;\; a} \pi_{i, \alpha} I^a,
\end{equation}
where $\{ I^a \}$ is the basis of $\g$ dual to $\{ I_a\}$ with respect to the bilinear form $\langle \cdot, \cdot \rangle$.
We therefore have a coordinate free description of the phase space, with the canonical Poisson brackets in local coordinates being equivalent to 
\begin{subequations} \label{X h brackets}
\begin{align}
\label{X h brackets a} \{h_{i\1},h_{j\2}\} &= 0,\\
\label{X h brackets b} \{X_{i\1}, h_{j\2} \} &= h_{i\2} C_{\1\2} \delta_{ij},\\
\label{X h brackets c} \{X_{i\1}, X_{j\2} \} &= - [C_{\1\2},X_{i\2}]\delta_{ij}.
\end{align}
\end{subequations}
for each $i, j = 1, \ldots, N$.

Using the definition of the matrix $L^a_{\;\;\alpha}$ in \eqref{L a alpha def} we have
\begin{equation*}
L^\alpha_{\;\; a} \langle u_i, h_i^{-1} \partial_\alpha h_i \rangle I^a = \langle u_i, L^\alpha_{\;\; a} L^b_{\;\; \alpha} I_b \rangle I^a = \langle u_i, I_a \rangle I^a = u_i.
\end{equation*}
It then follows from the expression \eqref{pi i alpha} for $\pi_{i, \alpha}$ above, derived from the Lagrangian, and the definition \eqref{Xi def} of $X_i$ that we have a primary constraint of the form
\begin{equation} \label{constraints Ci}
\mathcal C_i \coloneqq X_i + u_i \approx 0
\end{equation}
for each defect $i=1, \dots, N$. These $N$ primary constraints are not entirely first or second class. Indeed, their Poisson brackets 
\begin{equation} \label{Ci Cj bracket}
\{\mathcal{C}_{i\1},\mathcal{C}_{j\2}\} 
= 
\{X_{i\1},X_{j\2}\} = - [C_{\1\2},\mathcal C_{i\2} - u_{i\2}] \delta_{ij} \approx [C_{\1\2}, u_{i\2}] \delta_{ij},
\end{equation}
are non-vanishing on the constraint surface \eqref{constraints Ci} and are not generally invertible.

Let $\{ v^i_p \}_{p=1}^{d_i}$ be a basis of the centraliser $\g^{u_i} \coloneqq \ker(\textup{ad}_{u_i})$ of the element $u_i \in \g$, with $d_i \coloneqq \dim \g^{u_i}$ for each $i = 1, \ldots, N$.
The first class part of each $\mathcal C_i$ is given by the set of constraints $\mathcal C^p_i \coloneqq \langle v^i_p, \mathcal C_i\rangle$ for $p = 1, \ldots, d_i$. These satisfy the relations
\begin{equation} \label{Ci first class part}
\{ \mathcal{C}^p_i, \mathcal{C}_j \} \approx \big\langle v^i_{p\1}, [C_{\1\2}, u_{i\2}] \big\rangle_\1 \delta_{ij} = [v^i_p, u_i] \delta_{ij} = 0,
\end{equation}
for every $i, j = 1, \ldots, N$ and $p = 1, \ldots, d_i$, where the last equality uses the fact that $v^i_p \in \g^{u_i}$. In particular, we have $\{ \mathcal{C}^p_i, \mathcal{C}^q_j \} \approx 0$ for any $q=1, \ldots, d_j$ so that the set of constraints $\mathcal C^p_i$ for $p = 1, \ldots, d_i$, $i = 1, \ldots, N$ are indeed first class. It also follows from \eqref{X h brackets b} that the first class constraints $\mathcal C^p_i$ generate right multiplication of the $h_i$ by elements $e^{\epsilon v^i_p}$ of the centraliser $G^{u_i}$ of $u_i$ in $G$ \--- note that under such transformations the $\g$-valued variables $\widehat u_i$ are invariant.

Let us extend the basis $\{ v^i_p \}_{p=1}^{d_i}$ of the centraliser $\g^{u_i}$ to a basis $\{ v^i_p \}_{p=1}^{d_i} \cup \{ \widetilde{v}^i_r \}_{r=1}^{c_i}$ of $\g$ where $c_i \coloneqq \dim \g - d_i$. We claim that the remaining constraints $\widetilde{\mathcal C}^r_i \coloneqq \langle \widetilde{v}^i_r, \mathcal C_i\rangle$ for $r = 1, \ldots, c_i$ contained in $\mathcal C_i$ are second class. We need to show that the matrix $\{ \widetilde{\mathcal C}^r_i, \widetilde{\mathcal C}^s_i \}$ for $r,s = 1, \ldots, c_i$ is invertible on the contraint surface $\mathcal C_i \approx 0$. If this were not the case then we would have $\sum_{s=1}^{c_i} \{ \widetilde{\mathcal C}^r_i, \widetilde{\mathcal C}^s_i \} a_s \approx 0$ for some $a_s \in \CC$ with $s = 1, \ldots, c_i$. On the other hand, we also know from \eqref{Ci first class part} that $\sum_{s=1}^{c_i} \{ \mathcal C^p_i, \widetilde{\mathcal C}^s_i \} a_s \approx 0$ for all $p = 1, \ldots, d_i$. Combining these statements we have
\begin{align*}
0 \approx \sum_{s=1}^{c_i} \{ \mathcal C_i, \widetilde{\mathcal C}^s_i \} a_s = \sum_{s=1}^{c_i} \big\{ \mathcal C_i, \langle \widetilde v^i_s, \mathcal C_i \rangle \big\} a_s \approx \sum_{s=1}^{c_i} \big\langle \widetilde v^i_{s\2}, [C_{\1\2} , u_{i\2}] \big\rangle_{\2} a_s
= \bigg[ u_i, \sum_{s=1}^{c_i} a_s \widetilde v^i_s \bigg],
\end{align*}
where in the third step we used \eqref{Ci Cj bracket}.
It follows that $\sum_{s=1}^{c_i} a_s \widetilde v^i_s \in \g^{u_i}$ which contradicts the assumption that $\{ \widetilde{v}^i_r \}_{r=1}^{c_i}$ is the basis of some complement of $\g^{u_i}$ in $\g$.

We would like to impose suitable gauge fixing conditions $\mathcal D^p_i \approx 0$, for $p = 1, \ldots, d_i$, to fix the first class constraints $\mathcal C^p_i$ and move to a Dirac bracket $\{ \cdot, \cdot \}^\ast$ which fixes the constraints $\mathcal C_i \approx 0$ strongly. In particular, we would like to compute the Dirac bracket $\{ \widehat{u}_{i\1}, \widehat{u}_{j\2} \}^\ast$ of the $\g$-valued variables $\widehat{u}_i = h_i u_i h_i^{-1}$ for $i=1, \ldots, N$. It turns out that the result is independent of the choice of gauge fixing condition $\mathcal D^p_i \approx 0$. Indeed, consider the variables $\widehat{X}_i \coloneqq h_i X_i h_i^{-1}$. One deduces from \eqref{X h brackets} that they have the Poisson brackets
\begin{subequations} \label{hXhI brackets}
\begin{align}
\label{hXhI brackets a} \{ \widehat{X}_{i \1}, \widehat{X}_{j \2} \} &= [ C_{\1\2}, \widehat{X}_{i\2} ]\delta_{ij},\\
\label{hXhI brackets b} \{ X_{i\1}, \widehat{X}_{j \2} \} &=0
\end{align}
\end{subequations}
for each $i, j = 1, \ldots, N$.
In particular, it follows from \eqref{hXhI brackets b} that $\{ \mathcal C_{i\1}, \widehat{X}_{j\2} \} = 0$ for any $i,j = 1, \ldots, N$. Now the matrix of Poisson brackets between the set of all second class constraints $\mathcal C^p_i$, $\mathcal D^p_i$ for $p = 1, \ldots, d_i$ and $\widetilde{\mathcal C}^r_i$ for $r = 1, \ldots, c_i$ is of the block form
\begin{equation} \label{PB constraints}
\left(
\begin{matrix}
0 & \ast & 0\\
\ast & \ast & \ast\\
0 & \ast & \ast
\end{matrix}\right)
\end{equation}
where the first, second and third block rows and columns correspond to the set of constraints $\mathcal C^p_i$, $\mathcal D^p_i$ and $\widetilde{\mathcal C}^r_i$, respectively. Each `$\ast$' denotes a possibly non-zero block matrix. The matrix \eqref{PB constraints} is invertible since the blocks in position $(1,2)$, $(2,1)$ and $(3,3)$ are all invertible by design. Its inverse is then schematically of the block form
\begin{equation} \label{inverse PB constraints}
\left(
\begin{matrix}
0 & \ast & 0\\
\ast & \ast & \ast\\
0 & \ast & \ast
\end{matrix}\right)^{-1} = \left(
\begin{matrix}
\ast & \ast & \ast\\
\ast & 0 & 0\\
\ast & 0 & \ast
\end{matrix}\right).
\end{equation}
Since $\{ \mathcal C^p_i, \widehat{X}_j \} = \{ \widetilde{\mathcal C}^r_i, \widehat{X}_j \} = 0$ for all $p = 1, \ldots, d_i$ and $r = 1, \ldots, c_i$, the zero block in the middle of the right hand side of \eqref{inverse PB constraints} implies that the Poisson brackets \eqref{hXhI brackets a} will remain unchanged when passing to the Dirac bracket, \emph{i.e.} we have
\begin{equation*}
\{ \widehat{X}_{i \1}, \widehat{X}_{j \2}\}^\ast = [ C_{\1\2}, \widehat{X}_{i\2}]\delta_{ij}.
\end{equation*}
Finally, using the fact that $\widehat{X}_i = - \widehat{u}_i$ after imposing the constraint $\mathcal C_i = 0$ strongly, we deduce that the $\g$-valued variables $\widehat{u}_i$ for $i=1, \ldots, N$ satisfy $N$ commuting copies of the Kostant-Kirillov bracket
\begin{equation} \label{Kostant-Kirillov}
\{ \widehat{u}_{i \1}, \widehat{u}_{j \2}\}^\ast = - [ C_{\1\2}, \widehat{u}_{i\2}]\delta_{ij}.
\end{equation}
To avoid overburdening the notation, and since we shall need to introduce a further Dirac bracket in \S\ref{sec: gauge fixing}, we will denote the Dirac bracket $\{ \cdot, \cdot \}^\ast$ introduced above simply as $\{ \cdot, \cdot \}$ from now on.

\subsection{Hamiltonian and secondary constraints} \label{sec: Ham and second constraints}

The Hamiltonian density is defined as the Legendre transform of the Lagrangian density \eqref{Lagrangian density}. However, since the field $A_t$ is non-dynamical, \emph{i.e.} there are no time derivatives of $A_t$ in the action, we shall perform the Legendre transform only with respect to the dynamical fields $A_{\bar z}$, $B_z$ and the dynamical variables $h_i$. So we define
\begin{align*}
\mathcal H &\coloneqq \langle \Pi_{\bar z}, \partial_t A_{\bar z} \rangle + \langle P_z, \partial_t B_z \rangle + \langle X_i, h_i^{-1}\partial_t h_i \rangle - \mathcal L\big( A,B, \{h_i\}_{i=1}^N \big)
\\
&= \frac{1}{2 \pi \ii} \langle \mathcal C_z, \partial_t A_{\bar z} \rangle + \langle P_z, \partial_t B_z \rangle + \sum_{i=1}^N\langle \mathcal C_i, h_i^{-1}\partial_t h_i \rangle
\\
&\qquad\qquad\qquad - \frac{1}{2 \pi \ii} \langle B_z, \partial_{\bar z} A_t + [A_{\bar z},A_t] \rangle + \sum_{i=1}^N \langle \widehat u _i, A_t \rangle \delta_{zz_i} + H^P_w \delta_{zw}
\end{align*}
where in the second line we have used the definition of the bulk constraint $\mathcal C_z$ in \eqref{bulk constraints} and of the defect constraints $\mathcal C_i$ for $i = 1, \ldots, N$ in \eqref{constraints Ci}. Since we have already set these along with $P_z$ strongly to zero, we can drop the corresponding terms in the Hamiltonian density.

The Hamiltonian is the integral of the Hamiltonian density over $C$, namely
\begin{align*}
H &\coloneqq \int_C \mathcal H \, \d z \wedge \d\bar z\\
&= - \frac{1}{2 \pi \ii} \big\llangle B_z, \partial_{\bar z} A_t + [A_{\bar z},A_t] \big\rrangle + \int_C \bigg( \sum_{i=1}^N \langle \widehat u _i, A_t \rangle \delta_{zz_i} \bigg) \d z \wedge \d\bar z + H^P_w,
\end{align*}
where in the first term of the right hand side we introduced the notation
\begin{equation*}
\llangle X, Y \rrangle \coloneqq \int_C \langle X, Y \rangle \, \d z \wedge \d\bar z
\end{equation*}
for any $\g$-valued fields $X$ and $Y$ on $C$.

Let $\mu$ denote the moment map of the Hitchin system \cite{Hitchin} (we refer the reader to \cite[\S7.11]{BBT} for a concise review of Hitchin systems)
\begin{equation}
\mu \coloneqq \frac{1}{2\pi \ii} \big( \partial_{\bar z}B_z + \comm{A_{\bar z}}{B_z} \big) = -  \partial_{\bar z}\Pi_{\Bar{z}} - [A_{\bar z},\Pi_{\Bar{z}}],
\end{equation}
where in the second equality we have used the constraint $\mathcal C_z = 0$ in \eqref{bulk constraints} which is now imposed strongly.
Introducing also the $\g$-valued field
\begin{equation} \label{hat mu def}
\widehat \mu \coloneqq \mu + \sum_{i=1}^N \widehat u_i \delta_{zz_i},
\end{equation}
the Hamiltonian can be rewritten succinctly as
\begin{equation} \label{Hamiltonian with mu}
H = \llangle \widehat  \mu , A_t\rrangle + H^P_w.
\end{equation}

\subsubsection{Gauge invariance}

We need to ensure that the remaining primary constraint, $\Pi_t \approx 0$, is preserved under time evolution. That is,  
\begin{equation*}
\{H,\Pi_t\} = \widehat \mu \approx 0,
\end{equation*} 
giving rise to the secondary constraint \(\widehat \mu \approx 0\). We see from the canonical brackets \eqref{DB for Cz Pz} that $-\widehat \mu$ is the generator of gauge transformations \eqref{gauge transf A B} on the fields $A_{\bar z}$ and $B_z$ since
\begin{subequations} \label{PB hat mu A B}
\begin{align}
\label{PB hat mu A} \{\widehat \mu_\1(z), A_{\Bar{z}\2}(z') \} &= - [C_{\1\2}, A_{\Bar{z}\2}(z)]\delta_{zz'} - \partial_{\Bar{z}}( C_{\1\2} \delta_{zz'})\\
\label{PB hat mu B} \{\widehat \mu_\1(z), B_{z\2}(z')\} &=  - [C_{\1\2}, B_{z\2}(z)] \delta_{zz'}.
\end{align}
\end{subequations}
Note that the moment map \(\mu\) satisfies the following Poisson bracket
\begin{align*}
&\{\mu_\1(z), \mu_\2(z') \} =
\frac{1}{2\pi \ii}\{\mu_\1(z),\partial_{\bar{z}'}B_{z\2}(z') \} + \frac{1}{2\pi \ii}\big\{\mu_\1(z), [A_{\bar{z}\2}(z'),B_{z\2}(z')] \big\}\\
&\quad = \frac{1}{2 \pi \ii} \Big( -\partial_{\Bar{z}'}[C_{\1\2} \delta_{zz'},B_{z\2}(z')] - \big[ A_{\Bar{z}\2}(z), [C_{\1\2},B_{z\2}(z')] \big]\delta_{zz'}\\
&\qquad\qquad\qquad\qquad\qquad\qquad
- \big[ [C_{\1\2}, A_{\Bar{z}\2}(z)]\delta_{zz'}+ \partial_{\Bar{z}}(C_{\1\2} \delta_{zz'}), B_{z\2}(z') \big] \Big)\\
&\quad =\frac{1}{2\pi\ii}\Big( - [C_{\1\2},\partial_{\Bar{z}}B_{z\2}(z)] \delta_{zz'} - \big[ C_{\1\2},[A_{\Bar{z}\2}(z), B_{z\2}(z')] \big]\delta_{zz'} \Big) = - \comm{C_{\1\2}}{\mu_\2(z)} \delta_{zz'},
\end{align*}
where in the second equality we used the relations \eqref{PB hat mu A B}, which also trivially hold with $\widehat{\mu}$ replaced by $\mu$. In the third equality we have used the Jacobi identity and the fact that 
$\partial_{\bar{z}} \delta_{zz'} + \partial_{\bar{z}'} \delta_{zz'} = 0$, which follows using the identity $\partial_{\bar z} (z-z')^{-1} = - 2 \pi \ii \delta_{zz'}$.

The Poisson bracket of \(\widehat \mu\) with itself is therefore
\begin{align*}
\{\widehat\mu_\1(z), \widehat \mu_\2(z')\} &= \{\mu_\1(z),\mu_\2(z') \} + \sum_{i,j=1}^N 
\{ \widehat u_{i\1}, \widehat u_{j\2} \} \delta_{zz_i}\delta_{z'z_j}
\\
&= - [C_{\1\2}, \mu_\2(z)]\delta_{zz'} - \sum_{i=1}^N [ C_{\1\2}, \widehat u_{i\2} \delta_{zz_i} ] \delta_{z'z_i}
= - [C_{\1\2}, \widehat \mu_\2(z)] \delta_{zz'}
\end{align*}
where in the second equality we have used \eqref{Kostant-Kirillov} for the second term.
This vanishes on the constraint surface so $\widehat \mu$ is first class \--- we will set it strongly to zero with an appropriate gauge fixing condition in the following section.

The time evolution of \(\widehat \mu\) is given by
\begin{align*}
\{H, \widehat \mu(z) \} &\approx \frac{1}{2 \pi \ii} \big\{ H^P_w, [A_{\bar{z}}(z), B_z(z)] \big\}\\
&= \frac{1}{2 \pi \ii} \big[ \{ H^P_w, A_{\bar{z}}(z) \}, B_z(z) \big]
= - \comm{P'(B_z(w))}{B_z(z)}\delta_{zw} = 0,
\end{align*}
and therefore we have no tertiary constraints.

\subsection{Gauge fixing and Lax formalism} \label{sec: gauge fixing}

We wish to fix the gauge invariance arising from the constraint $\widehat\mu \approx 0$. We will use the gauge fixing condition $A_{\bar z} \approx 0$ and simultaneously impose this condition and the constraint $\widehat\mu \approx 0$ strongly by defining a new Dirac bracket. To this end, recall that 
\begin{equation*}
\{\widehat \mu_\1(z), A_{\bar{z}\2}(z')\} = - [C_{\1\2}, A_{\bar z\2}(z)] \delta_{zz'} - \partial_{\bar{z}} (C_{\1\2} \delta_{zz'}) \approx - \partial_{\bar{z}} (C_{\1\2} \delta_{zz'})
\end{equation*}
where the first equality is \eqref{PB hat mu A} and in the last step we have used the new constraint $A_{\bar z} \approx 0$. This can be inverted, since
\begin{equation*}
\left \llangle - \partial_{\bar{z}}( C_{\1\2} \delta_{zz'}), \frac{1}{2\pi\ii} \frac{C_{\2\3}}{z' - z''} \right \rrangle_{(z', \2)}
= \frac{\ii}{2\pi} C_{\1\3} \partial_{\bar{z}} \left( \frac{1}{z-z''} \right) =\frac{\ii}{2\pi} C_{\1\3}(-2\pi\ii \delta_{zz''})=  C_{\1\3}\delta_{zz''}.
\end{equation*}
Here the subscript $(z', \2)$ means that the pairing $\langle \cdot, \cdot \rangle$ is taken in the second tensor space and the integration is with respect to $z'$.
We therefore define the new Dirac bracket, denoted $\{\cdot, \cdot\}^\star$ for $\g$-valued functions $U$ and $V$ on $C$, by
\begin{align*}
&\{U_\1(z), V_\2(z')\}^\star = \{U_\1(z), V_\2(z')\} \\
&\qquad - \left\llangle \{U_\1(z), \widehat\mu_\3(z'')\}, \left\llangle \frac{1}{2 \pi \ii} \frac{C_{\3\4}}{z'' - z'''},
\{A_{\bar{z}\4}(z'''),V_\2(z')\} \right\rrangle_{(z''',\4)} \right\rrangle_{(z'',\3)} \\
&\qquad - \left\llangle \{U_\1(z), A_{\bar{z}\3}(z''')\}, \left\llangle \frac{1}{2\pi \ii} \frac{C_{\3\4}}{z'' - z'''}, \{\widehat\mu_\4(z'''),V_\2(z')\} \right\rrangle_{(z''',\4)} \right\rrangle_{(z'',\3)}.
\end{align*}

\subsubsection{Lax algebra} \label{sec: Lax algebra}

We will show that the Dirac bracket of $B_z$ with itself satisfies the Lax algebra
\begin{equation} \label{Lax algebra}
\{B_{z\1}(z), B_{z\2}(z')\}^\star = \big[ r_{\1\2}(z,z'), B_{z\1}(z) + B_{z\2}(z') \big],
\end{equation}
where $r_{\1\2}(z,z')$ is the standard classical $r$-matrix
\begin{equation} \label{r-matrix}
r_{\1\2}(z,z') = \frac{C_{\1\2}}{z'-z}.
\end{equation}

To compute this Dirac bracket, we begin by noting that \eqref{PB hat mu B} implies 
\begin{equation*}
\{B_{z\1}(z), \widehat \mu_\2(z')\} = [C_{\1\2}, B_{z\1}(z)] \delta_{zz'}.
\end{equation*}
Using this and the bracket $\{ B_{z \1}(z), A_{\bar z \2}(z') \} = -2 \pi \ii C_{\1\2} \delta_{zz'}$ which follows from \eqref{DB for Cz Pz b} along with the constraint $\mathcal C_z = 0$ in \eqref{bulk constraints}, we find
\begin{align*}
&\{B_{z\1}(z), B_{z\2}(z')\}^\star\\
&\qquad = - \left\llangle [C_{\1\3},B_{z\1}(z)] \delta_{zz''}, \left\llangle \frac{1}{2 \pi \ii} \frac{C_{\3\4}}{z'' - z'''}, 2 \pi \ii C_{\2\4}\delta_{z'z'''} \right\rrangle_{(z''',\4)} \right\rrangle_{(z'',\3)}\\
&\qquad\qquad -
\left\llangle - 2\pi \ii C_{\1\3}\delta_{zz''},\left\llangle 
\frac{1}{2 \pi \ii}\frac{C_{\3\4}}{z'' - z'''}, - [C_{\2\4}, B_{z\2}(z')] \delta_{z'z'''} \right\rrangle_{(z''',\4)} \right\rrangle_{(z'',\3)}\\
&\qquad 
= - \left \llangle \comm{C_{\1\3}}{B_{z\1}(z)} \delta_{zz''}, \frac{ C_{\2\3}}{z''-z'} \right \rrangle_{(z'',\3)} - \left \llangle  C_{\1\3}\delta_{zz''}, \comm{ \frac{C_{\2\3}}{z'' - z'}}{B_{z\2}(z')} \right \rrangle_{(z'',\3)}\\
&\qquad = - \comm{\frac{C_{\1\2}}{z-z'}}{B_{z\1}(z)} - \comm{\frac{C_{\1\2}}{z-z'}}{B_{z\2}(z')} =  \comm{\frac{C_{\1\2}}{z'-z}}{B_{z\1}(z) + B_{z\2}(z')}.
\end{align*}
In other words, we recover the Lax algebra \eqref{Lax algebra}.

\subsubsection{Lax matrix}

By definition of $\widehat\mu$ in \eqref{hat mu def}, it follows that setting this constraint and its gauge fixing condition to zero strongly, \emph{i.e.} $\widehat\mu = 0$ and $A_{\bar z} = 0$, leads to the equation
\begin{equation}
\partial_{\bar z}B_z = - 2\pi \ii \, \sum_{i=1}^N \widehat u_i \delta_{zz_i},
\end{equation}
or $\partial_{\bar z}B_z = - 2\pi \ii \, \widehat u_i \delta_{zz_i}$ in a small neighbourhood of the point $z_i$, which is equivalent to \eqref{Bz mero}. This then leads to the local meromorphic behaviour \eqref{B mero Hitchin} of the $(1,0)$-form $B$, namely
\begin{equation*}
B = \frac{\widehat{u}_i}{z-z_i} \d z + O(1).
\end{equation*}
The Kostant-Kirillov bracket \eqref{Kostant-Kirillov} for the residues $\widehat{u}_i$ obtained in \S\ref{sec: defect can var} (recall that we are now denoting the Dirac bracket $\{\cdot, \cdot\}^\ast$ of \S\ref{sec: defect can var} simply as $\{\cdot, \cdot\}$) is equivalent to the Lax algebra \eqref{Lax algebra} derived in \S\ref{sec: Lax algebra}.

\subsubsection{Lax equation}

At this point we have now fixed all the constraints strongly except for the primary constraint $\Pi_t \approx 0$. However, now that $\widehat\mu = 0$ is imposed strongly, the Hamiltonian \eqref{Hamiltonian with mu} no longer involves the field $A_t$ and simply reduces to
\begin{equation*}
H = H^P_w.
\end{equation*}
In particular, together with the Dirac bracket \eqref{Lax algebra} this now implies the Lax equation \eqref{Lax equation Bz} in the Hamiltonian formalism
\begin{equation} \label{Lax equation Bz Ham}
\big\{ H^P_w, B_z(z) \big\}^\star = \bigg[ \frac{P'\big( B_z(w) \big)}{z - w}, B_z(z) \bigg].
\end{equation}
We deduce, as claimed at the end of \S\ref{sec: Lag analysis}, that the time flow $\partial_t$ along the topological direction of the $3$-dimensional space $\RR \times C$ is the one induced by the Hamiltonian $H^P_w = P(B_z(w))$ with respect to the Dirac bracket, \emph{i.e.} $\partial_t f = \{ H^P_w, f \}^\star$ for any function $f$ of the Lax matrix $B_z$. Focusing on such observables, we are also free to set $\Pi_t = 0$ strongly since these all Poisson commute with $\Pi_t$ under the Dirac bracket $\{ \cdot, \cdot \}^\star$ and so their bracket will remain unchanged after introducing a further Dirac bracket to fix the constraint $\Pi_t \approx 0$.

\subsubsection{Involution}

It is well know that the Lax algebra \eqref{Lax algebra} implies the involution property
\begin{equation} \label{involution DB}
\big\{ H^P_w, H^Q_z \big\}^\star = 0,
\end{equation}
for any pair of $G$-invariant polynomials $P, Q : \g \to \CC$ and distinct points $w,z \in C$.

This can also be seen more directly from the above Hamiltonian analysis of $3$d mixed BF theory as follows. Since $H^P_w = P(B_z(w))$ only depends on the field $B_z$ we have the involution property
\begin{equation} \label{involution PB}
\big\{ H^P_w, H^Q_z \big\} = 0
\end{equation}
with respect to the Poisson bracket (more precisely, recall that $\{ \cdot, \cdot\}$ denotes the Dirac bracket introduced in \S\ref{sec: primary constraints}), for \emph{any} polynomials $P, Q : \g \to \CC$ and distinct points $w, z \in C$. But since $H^P_w$ is gauge invariant for \emph{$G$-invariant} polynomials $P$ and $- \widehat{\mu}$ is the generator of gauge transformations, see \eqref{PB hat mu A B}, we have $\{\widehat{\mu}(z), H^P_w\} = 0$. The involution property \eqref{involution PB}, for any polynomials $P, Q : \g \to \CC$, therefore immediately implies the involution property \eqref{involution DB}, for any $G$-invariant polynomials $P, Q : \g \to \CC$.

\section{Discussion} \label{sec: discussion}

In this article we showed that Gaudin models associated with a finite-dimensional semisimple Lie algebra, and more generally tamely ramified Hitchin systems, can be obtained from $3$d mixed BF theory in the presence of certain line defects by moving to the Hamiltonian framework and fixing the gauge symmetry using the gauge fixing condition $A_{\bar z} \approx 0$.

This $3$-dimensional gauge theoretic origin of finite-dimensional Gaudin models is exactly analogous to that of affine Gaudin models in terms of $4$-dimensional mixed topological-holomorphic Chern-Simons theory \cite{Vicedo:2019dej}.

\subsection{Alternative realisations}

The Lax matrix of the Gaudin model, or the Higgs field of the Hitchin system, arises from the $(1,0)$-form $B$ of the $3$d mixed BF theory. In particular, after going to the gauge $A_{\bar z} = 0$ the latter becomes meromorphic with simple poles \eqref{B mero Hitchin} at each $z_i$, the location of the type $A$ line defects. The specific choice of line defect \eqref{A defect} led to the residues of $B$ at these simple poles being coadjoint orbits $\widehat{u}_i = h_i u_i h_i^{-1}$ of some fixed Lie algebra elements $u_i \in \g$. As is well known, and as we have rederived in the present setting in \S\ref{sec: defect can var}, such coadjoint orbits provide a realisation of the Kostant-Kirillov Poisson bracket \eqref{Kostant-Kirillov}.

It would be interesting to see if other realisations of the Kostant-Kirillov Poisson bracket can be obtained by making other choices of type $A$ defects than \eqref{A defect}. Indeed, since the field $B_z$ satisfies the Lax algebra \eqref{Lax algebra} regardless of the choice of type $A$ line defects we make, the residues $\widehat u_i$ at each simple pole $z_i$ of $B_z$ will necessarily satisfy the Kostant-Kirillov bracket. As mentioned in the affine case in \cite[\S4.1]{Vicedo:2019dej}, it would be desirable to find the precise dictionary between the possible choices of type $A$ line defects one can introduce in $3$d mixed BF theory and the different types of possible representations of the Kostant-Kirillov bracket.

\subsection{Generalised Gaudin models}

We have focused in this paper on the case when the Lax matrix of the Gaudin model, or the Higgs field of the Hitchin system, has simple poles at the marked points $z_i$.

It would be interesting to consider also type $A$ line defects which would give rise to higher order poles in the Lax matrix in order to construct Gaudin models with irregular singularities \cite{FFT, FFRy, VY2}. In the affine setting, generalised surface defects in $4$-dimensional Chern-Simons theory leading to affine Gaudin models with irregular singularities were recently considered in \cite{Benini:2020skc, Lacroix:2020flf}.

Other generalisations of the Gaudin model which one could try to relate to $3$d mixed BF theory, or some generalisation thereof, include cyclotomic Gaudin models \cite{Skrypnyk, Vicedo:2014zza, VY2} or dihedral Gaudin models (see \cite{Vicedo:2017cge} in the affine case), whose Lax matrices are equivariant under the action of cyclic or dihedral groups, respectively. In the affine case, such a generalisation was considered recently in \cite{Schmidtt:2020dbf} where the symmetric space $\lambda$-model, which can be described as a $\ZZ_4$-cyclotomic affine Gaudin model, was obtained along the lines of \cite{Vicedo:2019dej} starting from $4$d Chern-Simons theory with a $\ZZ_4$-equivariance condition imposed on the gauge field.

\subsection{Quantum Gaudin models}

The $4$-dimensional gauge theoretic origin of $2$-dimensional integrable field theories, as proposed by Costello and Yamazaki in \cite{Costello:2019tri}, has been extensively studied over the past couple of years, see for instance \cite{Delduc:2019whp, Schmidtt:2019otc, Fukushima:2020kta, Fukushima:2020dcp, Costello:2020lpi, Tian:2020ryu, Tian:2020pub, Benini:2020skc, Bittleston:2020hfv, Lacroix:2020flf, Caudrelier:2020xtn, Fukushima:2020tqv, Affleck:2021ypq, Fukushima:2021eni, Derryberry:2021rne, Stedman:2021wrw, Schmidtt:2020dbf, Lacroix:2021iit, Fukushima:2021ako}.

The proposal of \cite{Vicedo:2017cge}, see also \cite{Delduc:2019bcl, Lacroix:2018njs}, to reformulate non-ultralocal integrable field theories with twist functions as affine Gaudin models similarly provides a deeper origin, more algebraic in nature, of the integrable structure in these theories.

\medskip

Both the gauge theoretic and algebraic approaches to $2$-dimensional integrable field theories, which are of course intimately related \cite{Vicedo:2019dej}, have been used to construct many new examples of $2$-dimensional classical integrable field theories in recent years; see for instance \cite{Delduc:2018hty, Delduc:2019bcl, Bassi:2019aaf, Arutyunov:2020sdo} in the affine Gaudin model setting and the references above in the $4$d Chern-Simons theory setting. Finite Gaudin models, or equivalently $3$d mixed BF theory, could similarly be used to extend the list of known finite-dimensional integrable systems.

\medskip

However, the main interest in both approaches lies in their potential to offer new perspectives on various long-standing open problems in \emph{quantum} integrable field theory, such as the problem of quantisation of non-ultralocal integrable field theories or the search for a deeper understanding of the celebrated ODE/IM correspondence \cite{Dorey:1998pt, Bazhanov:1998wj, Bazhanov:2003ni, Feigin:2007mr, Lukyanov:2010rn}. Indeed, one of the main original motivations in \cite{Vicedo:2017cge} for reformulating non-ultralocal integrable field theories as affine Gaudin models was the remarkable observation made in \cite{Feigin:2007mr}, based on the example of quantum KdV theory, that this may provide an explanation for the ODE/IM correspondence in terms of some suitable affine generalisation of the geometric Langlands correspondence.

By contrast with the affine case, however, the quantisation of the finite Gaudin model, and more generally of the Hitchin system, is extremely well understood; see \emph{e.g.} \cite{BD91, FFR, Frenkel1, Frenkel2, MV1, MV2, MV3, MV4, MTV1, FFT, FFRy, Rybnikov}. The connection between $3$d mixed BF theory and finite Gaudin models should therefore provide a useful toy model for further developing our understanding of the gauge theoretic approach to integrable models and more generally integrable field theories in the sense of \cite{Costello:2019tri}. In particular, it would be very desirable to understand the Bethe ansatz construction in quantum Gaudin models, and more generally the Gaudin/oper correspondence \cite{Frenkel1, Frenkel2, MV1, MV2, FFT, FFRy}, from the point of view of quantum $3$d mixed BF theory.


\begin{thebibliography}{99}

\bibitem[ABW]{Affleck:2021ypq}
I.~Affleck, D.~Bykov and K.~Wamer,
\emph{Flag manifold sigma models: spin chains and integrable theories},
arXiv:2101.11638 [hep-th].

\bibitem[ABL]{Arutyunov:2020sdo}
G.~Arutyunov, C.~Bassi and S.~Lacroix,
\emph{New integrable coset sigma models},
JHEP \textbf{03} (2021), 062.

\bibitem[BBT]{BBT}
O.~Babelon, D.~Bernard, M.~Talon,
\emph{Introduction to classical integrable systems},
2003, Cambridge University Press.

\bibitem[BL]{Bassi:2019aaf}
C.~Bassi and S.~Lacroix,
\emph{Integrable deformations of coupled $\sigma$-models},
JHEP \textbf{05} (2020), 059.

\bibitem[BLZ1]{Bazhanov:1998wj}
V.~V.~Bazhanov, S.~L.~Lukyanov and A.~B.~Zamolodchikov,
\emph{Spectral determinants for Schrodinger equation and Q operators of conformal field theory},
J. Statist. Phys. \textbf{102} (2001), 567-576.

\bibitem[BLZ2]{Bazhanov:2003ni}
V.~V.~Bazhanov, S.~L.~Lukyanov and A.~B.~Zamolodchikov,
\emph{Higher level eigenvalues of Q operators and Schroedinger equation},
Adv. Theor. Math. Phys. \textbf{7} (2003) no.4, 711-725.

\bibitem[BD]{BD91}
A.~Beilinson and V.~Drinfel'd, \emph{Quantization of Hitchin's integrable
system and Hecke eigensheaves}, available at
\url{www.math.uchicago.edu/~mitya/langlands/hitchin/BD-hitchin.pdf}.

\bibitem[BSV]{Benini:2020skc}
M.~Benini, A.~Schenkel and B.~Vicedo,
\emph{Homotopical analysis of 4d Chern-Simons theory and integrable field theories},
arXiv:2008.01829 [hep-th].

\bibitem[BS]{Bittleston:2020hfv}
R.~Bittleston and D.~Skinner,
\emph{Twistors, the ASD Yang-Mills equations, and 4d Chern-Simons theory},
arXiv:2011.04638 [hep-th].

\bibitem[CSV]{Caudrelier:2020xtn}
V.~Caudrelier, M.~Stoppato and B.~Vicedo,
\emph{On the Zakharov-Mikhailov action: $4$d Chern-Simons origin and covariant Poisson algebra of the Lax connection},
Lett. Math. Phys. \textbf{111} (2021), 82.

\bibitem[C1]{Costello:2013zra}
K.~Costello,
\emph{Supersymmetric gauge theory and the Yangian},
arXiv:1303.2632. 

\bibitem[C2]{Costello:2013sla}
K.~Costello,
\emph{Integrable lattice models from four-dimensional field theories},
Proc.\ Symp.\ Pure Math.\  {\bf 88} (2014) 3,
arXiv:1308.0370.

\bibitem[CS]{Costello:2020lpi}
K.~Costello and B.~Stefa\'nski,
\emph{Chern-Simons Origin of Superstring Integrability},
Phys. Rev. Lett. \textbf{125} (2020) no.12, 121602.

\bibitem[CWY1]{Costello:2017dso}
K.~Costello, E.~Witten and M.~Yamazaki,
\emph{Gauge Theory and Integrability, I},
ICCM Not.\ {\bf 6} (2018) 46--119,
arXiv:1709.09993. 

\bibitem[CWY2]{Costello:2018gyb}
K.~Costello, E.~Witten and M.~Yamazaki,
\emph{Gauge Theory and Integrability, II},
ICCM Not.\ {\bf 6} (2018) 120--149,
arXiv:1802.01579. 

\bibitem[CY]{Costello:2019tri}
K.~Costello and M.~Yamazaki,
\emph{Gauge Theory And Integrability, III},

arXiv:1908.02289.

\bibitem[DLMV1]{Delduc:2018hty}
F.~Delduc, S.~Lacroix, M.~Magro and B.~Vicedo,
\emph{Integrable Coupled $\sigma$ Models},
Phys. Rev. Lett. \textbf{122} (2019) no.4, 041601.

\bibitem[DLMV2]{Delduc:2019bcl}
F.~Delduc, S.~Lacroix, M.~Magro and B.~Vicedo,
\emph{Assembling integrable $\sigma$-models as affine Gaudin models},
JHEP \textbf{06} (2019), 017.

\bibitem[DLMV3]{Delduc:2019whp}
F.~Delduc, S.~Lacroix, M.~Magro and B.~Vicedo,
\emph{A unifying 2d action for integrable $\sigma$-models from 4d Chern-Simons theory},
Lett. Math. Phys. \textbf{110} (2020), 1645-1687

\bibitem[D]{Derryberry:2021rne}
R.~Derryberry,
\emph{Lax formulation for harmonic maps to a moduli of bundles},
arXiv:2106.09781 [math.AG].

\bibitem[DT]{Dorey:1998pt}
P.~Dorey and R.~Tateo,
\emph{Anharmonic oscillators, the thermodynamic Bethe ansatz, and nonlinear integral equations},
J. Phys. A \textbf{32} (1999), L419-L425.

\bibitem[FFr]{Feigin:2007mr}
B.~Feigin and E.~Frenkel,
\emph{Quantization of soliton systems and Langlands duality},
Adv. Stud. Pure Math. {\bf 61}, Math. Soc. Japan, Tokyo, 2011.

\bibitem[FFR]{FFR}
B.~Feigin, E.~Frenkel, and N.~Reshetikhin,
\emph{Gaudin model, Bethe ansatz and correlation functions at the critical level},
Comm. Math. Phys. \textbf{166} (1994), 27--62.

\bibitem[FFRy]{FFRy}
B.~Feigin, E.~Frenkel, and L.~Rybnikov,
\emph{Opers with irregular singularity and spectra of the shift of argument subalgebra},
Duke Math. J. {\bf 155} (2010), no. 2, 337--363.

\bibitem[FFT]{FFT}
B.~Feigin, E.~Frenkel and V.~Toledano Laredo,
\emph{Gaudin models with irregular singularities},
Adv. Math. {\bf 223} (2010) 873.
  
\bibitem[Fr1]{Frenkel1}
E.~Frenkel,
\emph{Opers on the projective line, flag manifolds and Bethe Ansatz},
Mosc. Math. J. \textbf{4} (2004), no.~3, 655--705.

\bibitem[Fr2]{Frenkel2}
E.~Frenkel, 
\emph{Gaudin model and opers},
Progr. Math. \textbf{237} (2005), 1--58.

\bibitem[FSY1]{Fukushima:2020kta}
O.~Fukushima, J.~i.~Sakamoto and K.~Yoshida,
\emph{Comments on $\eta$-deformed principal chiral model from 4D Chern-Simons theory},
Nucl. Phys. B \textbf{957} (2020), 115080.

\bibitem[FSY2]{Fukushima:2020dcp}
O.~Fukushima, J.~i.~Sakamoto and K.~Yoshida,
\emph{Yang-Baxter deformations of the AdS$_5\times$S$^5$ supercoset sigma model from 4D Chern-Simons theory},
JHEP \textbf{09} (2020), 100.

\bibitem[FSY3]{Fukushima:2020tqv}
O.~Fukushima, J.~I.~Sakamoto and K.~Yoshida,
\emph{Faddeev-Reshetikhin model from a 4D Chern-Simons theory},
JHEP \textbf{02} (2021), 115.

\bibitem[FSY4]{Fukushima:2021eni}
O.~Fukushima, J.~i.~Sakamoto and K.~Yoshida,
\emph{Integrable deformed T$^{1,1}$ sigma models from 4D Chern-Simons theory},
JHEP \textbf{09} (2021), 037.

\bibitem[FSY5]{Fukushima:2021ako}
O.~Fukushima, J.~i.~Sakamoto and K.~Yoshida,
\emph{Non-Abelian Toda field theories from a 4D Chern-Simons theory},
arXiv:2112.11276 [hep-th].

\bibitem[G]{Gaudin}
M.~Gaudin,
\emph{Diagonalisation d’une classe d’Hamiltoniens de spins},
J. Physique 37 (1976), 1087–1098.

\bibitem[GRW]{GRW}
O.~Gwilliam, E.~Rabinovich and B.~R.~Williams,
\emph{Quantization of topological-holomorphic field theories: local aspects},
arXiv: 2107.06734 [math-ph].

\bibitem[GW]{GW}
O.~Gwilliam and B.~R.~Williams,
\emph{A one-loop exact quantization of ChernSimons theory},
arXiv: 1910.05230 [math-ph].

\bibitem[H]{Hitchin}
N.~Hitchin,
\emph{Stable bundles and integrable systems},
Duke Math. J. {\bf 54} (1) 91–114.

\bibitem[L1]{Lacroix:2018njs}
S.~Lacroix,
\emph{Integrable models with twist function and affine Gaudin models},
PhD thesis, arXiv:1809.06811 [hep-th].

\bibitem[L2]{Lacroix:2021iit}
S.~Lacroix,
\emph{4-dimensional Chern-Simons theory and integrable field theories},
arXiv:2109.14278 [hep-th].

\bibitem[LV]{Lacroix:2020flf}
S.~Lacroix and B.~Vicedo,
\emph{Integrable $\mathcal{E}$-Models, 4d Chern-Simons Theory and Affine Gaudin Models. I.~Lagrangian Aspects},
SIGMA \textbf{17} (2021), 058.

\bibitem[LZ]{Lukyanov:2010rn}
S.~L.~Lukyanov and A.~B.~Zamolodchikov,
\emph{Quantum Sine(h)-Gordon Model and Classical Integrable Equations},
JHEP \textbf{07} (2010), 008.

\bibitem[MV1]{MV1}
E.~Mukhin and A.~Varchenko,
\emph{Critical points of master functions and flag varieties},
Comm. Cont. Math. {\bf 6} (2004), no. 1, 111–163.

\bibitem[MV2]{MV2}
E.~Mukhin and A.~Varchenko,
\emph{Miura Opers and Critical Points of Master Functions},
Cent. Eur. J. Math. {\bf 3} (2005), 155–182.

\bibitem[MV3]{MV3}
E.~Mukhin and A.~Varchenko,
\emph{Norm of a Bethe vector and the Hessian of the master function},
Compositio Math. {\bf 141} (2005), 1012–1028.

\bibitem[MV4]{MV4}
E.~Mukhin, A.~Varchenko,
\emph{Multiple orthogonal polynomials and a counterexample to Gaudin Bethe Ansatz Conjecture},
Trans. Amer. Math. Soc. {\bf 359} (2007), no. 11, 5383-5418.

\bibitem[MTV1]{MTV1}
E.~Mukhin, V.~Tarasov, A.~Varchenko,
\emph{Schubert calculus and representations of the general linear group},
J. Amer. Math. Soc. {\bf 22} (2009), no. 4, 909-940.

\bibitem[Ra]{RaPhD}
E.~Rabinovich,
\emph{Factorization Algebras for Bulk-Boundary Systems},
PhD thesis, arXiv:2111.01757 [math.QA].

\bibitem[Ry]{Rybnikov}
L.~Rybnikov,
\emph{A proof of the Gaudin Bethe Ansatz conjecture},
Int. Math. Res. Not. 2020 (22), 8766-8785.

\bibitem[Sc1]{Schmidtt:2019otc}
D.~M.~Schmidtt,
\emph{Holomorphic Chern-Simons theory and lambda models: PCM case},
JHEP \textbf{04} (2020), 060.

\bibitem[Sc2]{Schmidtt:2020dbf}
D.~M.~Schmidtt,
\emph{Symmetric space \ensuremath{\lambda}-model exchange algebra from 4d holomorphic Chern-Simons theory},
JHEP \textbf{21} (2020), 004.

\bibitem[Sk]{Skrypnyk}
T.~Skrypnyk,
\emph{$\ZZ_2$-graded Gaudin models and analytical Bethe ansatz},
Nucl. Phys. B {\bf 870} (2013), no. 3, 495–529.

\bibitem[St]{Stedman:2021wrw}
J.~Stedman,
\emph{Four-Dimensional Chern-Simons and Gauged Sigma Models},
arXiv:2109.08101 [hep-th].

\bibitem[T]{Tian:2020ryu}
J.~Tian,
\emph{Comments on $\lambda$--deformed models from 4D Chern-Simons theory},
arXiv:2005.14554 [hep-th].

\bibitem[THC]{Tian:2020pub}
J.~Tian, Y.~J.~He and B.~Chen,
\emph{$\lambda$-Deformed $AdS_5 \times S^5$ superstring from 4D Chern-Simons theory},
Nucl. Phys. B \textbf{972} (2021), 115545.

\bibitem[VY1]{Vicedo:2014zza}
B.~Vicedo and C.~A.~S.~Young,
\emph{Cyclotomic Gaudin models: construction and Bethe ansatz},
Commun. Math. Phys. \textbf{343} (2016) no.3, 971--1024.

\bibitem[VY2]{VY2}
B.~Vicedo, C.~A.~S.~Young,
\emph{Cyclotomic Gaudin models with irregular singularities},
J. Geom. Phys. {\bf 121} (2017) 247-278.

\bibitem[V1]{Vicedo:2017cge}
B.~Vicedo,
\emph{On integrable field theories as dihedral affine Gaudin models},
Int. Math. Res. Not. \textbf{2020} (2020) no.15, 4513-4601

\bibitem[V2]{Vicedo:2019dej}
B.~Vicedo,
\emph{4d Chern–Simons theory and affine Gaudin models},
Lett. Math. Phys. {\bf 111}, 24 (2021).

\end{thebibliography}
\end{document}